\documentclass[apj,twocolumn,floatfix]{openjournal}

\usepackage{xcolor}
\usepackage{textgreek}
\usepackage[utf8]{inputenc}
\usepackage[english]{babel}

\usepackage{hyperref}
\hypersetup{
    unicode, 
    colorlinks=true,
    linkcolor=linkcolor,
    citecolor=linkcolor,
    filecolor=linkcolor,
    urlcolor=linkcolor,
}
\usepackage{color,colortbl}
\definecolor{linkcolor}{rgb}{0.0,0.3,0.5}
\usepackage{tensind}
\tensordelimiter{?}
\DeclareGraphicsExtensions{.bmp,.png,.jpg,.pdf}
\usepackage{verbatim}
\usepackage[normalem]{ulem}
\usepackage{orcidlink}

\usepackage{amsmath}

\graphicspath{{./}{figures/}}

\begin{document}

\title{JCMT Constraints on the Early-Time HCN and CO Emission and HCN Temporal Evolution of 3I/ATLAS}

\title{JCMT C\lowercase{onstraints} \lowercase{on} \lowercase{the} E\lowercase{arly}-T\lowercase{ime} HCN \lowercase{and} CO E\lowercase{mission} \lowercase{and} HCN T\lowercase{emporal} E\lowercase{volution} \lowercase{of} 3I/ATLAS}

\author{\vspace{-1.3cm}Jason~T.~Hinkle$^{1,2,3,*}$\orcidlink{0000-0001-9668-2920}}
\author{Bin~Yang$^{4,5}$\orcidlink{0000-0002-5033-9593}}
\author{Karen~J.~Meech$^{3}$\orcidlink{0000-0002-2058-5670}}
\author{Andrew~Hoffman$^{3}$\orcidlink{0000-0002-8732-6980}} 
\author{Benjamin~J.~Shappee$^{3}$\orcidlink{0000-0003-4631-1149}}
\author{W.~B.~Hoogendam$^{3\dagger}$\orcidlink{0000-0003-3953-9532}}
\author{James~J.~Wray$^{6,3}$\orcidlink{0000-0001-5559-2179}}

\affiliation{$^{1}$Department of Astronomy, University of Illinois Urbana-Champaign, 1002 West Green Street, Urbana, IL 61801, USA}
\affiliation{$^{2}$NSF-Simons AI Institute for the Sky (SkAI), 172 E. Chestnut St., Chicago, IL 60611, USA}
\affiliation{$^{3}$Institute for Astronomy, University of Hawai`i, 2680 Woodlawn Drive, Honolulu, HI 96822, USA}
\affiliation{$^{4}$Instituto de Estudios Astrof\'isicos, Facultad de Ingenier\'ia y Ciencias, Universidad Diego Portales, Santiago, Chile}
\affiliation{$^{5}$Planetary Science Institute, 1700 E Fort Lowell Rd STE 106, Tucson, AZ 85719, USA}
\affiliation{$^{6}$School of Earth and Atmospheric Sciences, Georgia Institute of Technology, 311 Ferst Drive, Atlanta, GA 30332, USA}

\altaffiltext{*}{NHFP Einstein Fellow}
\altaffiltext{\dag}{NSF Fellow}

\email{jhinkle6@illinois.edu}

\begin{abstract}

\noindent Interstellar objects (ISOs), particularly those with cometary activity, provide unique insight into the primordial physical and chemical conditions present during the formation of the planetary system in which they originated. Observations in the sub-mm regime allow for direct measurements of several parent molecules released from the comet nucleus into the coma. Here we present observations of the third ISO, 3I/ATLAS, with the `\=U`\=u heterodyne receiver on the James Clerk Maxwell Telescope (JCMT), which targeted emission from HCN($J = 3 - 2$) and CO($J = 2 - 1$). Our observations, taken between 16 July 2025 and 21 July 2025 (UT), when 3I/ATLAS was at a heliocentric distance between 4.01 and 3.84 au, provide the earliest sub-mm constraints on its activity. We do not detect HCN or CO in these epochs, with 3$\sigma$ upper-limits on the production rates of $Q(HCN) < 1.7 \times 10^{24}$ s$^{-1}$ at $r_h = 4.01\mbox{ -- }3.97$ au and $Q(CO) < 1.1 \times 10^{27}$ s$^{-1}$ at $r_h = 3.94\mbox{ -- }3.84$ au, respectively. We combine this HCN limit with later JCMT observations of HCN to constrain its temporal evolution. Fitting the HCN detections with a $Q(HCN) \propto r_h^{-n}$ model and accounting for the upper-limits yields $n = 12.7^{+6.9}_{-2.5}$. This slope is steeper than those of typical Solar System comets, but consistent with the production rate slopes measured for other species in the coma of 3I/ATLAS.
\end{abstract}


\section{Introduction} \label{sec:intro}

\begin{figure*}[t]
\centering
 \includegraphics[height=0.37\textwidth]{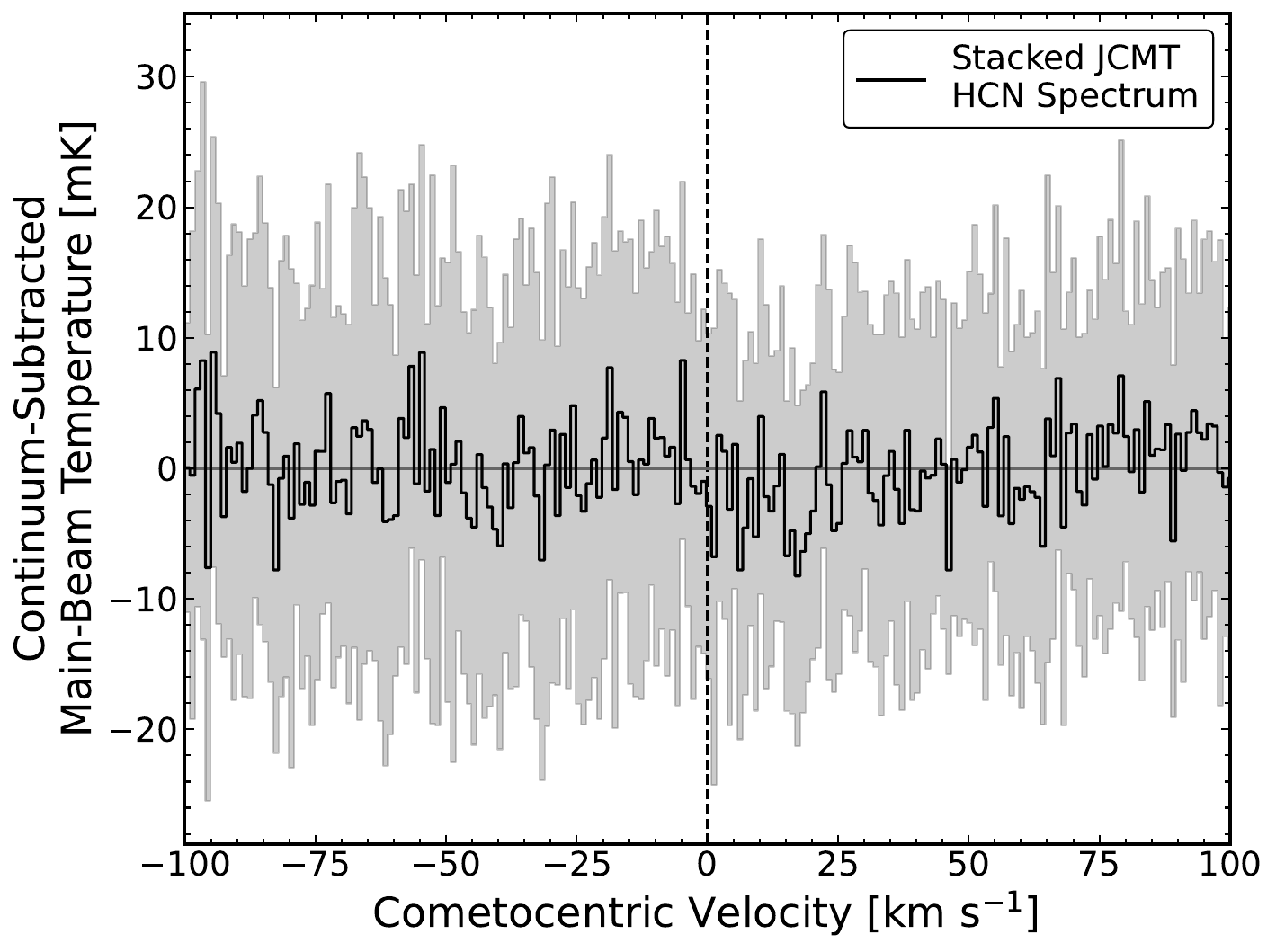}
 \includegraphics[height=0.37\textwidth]{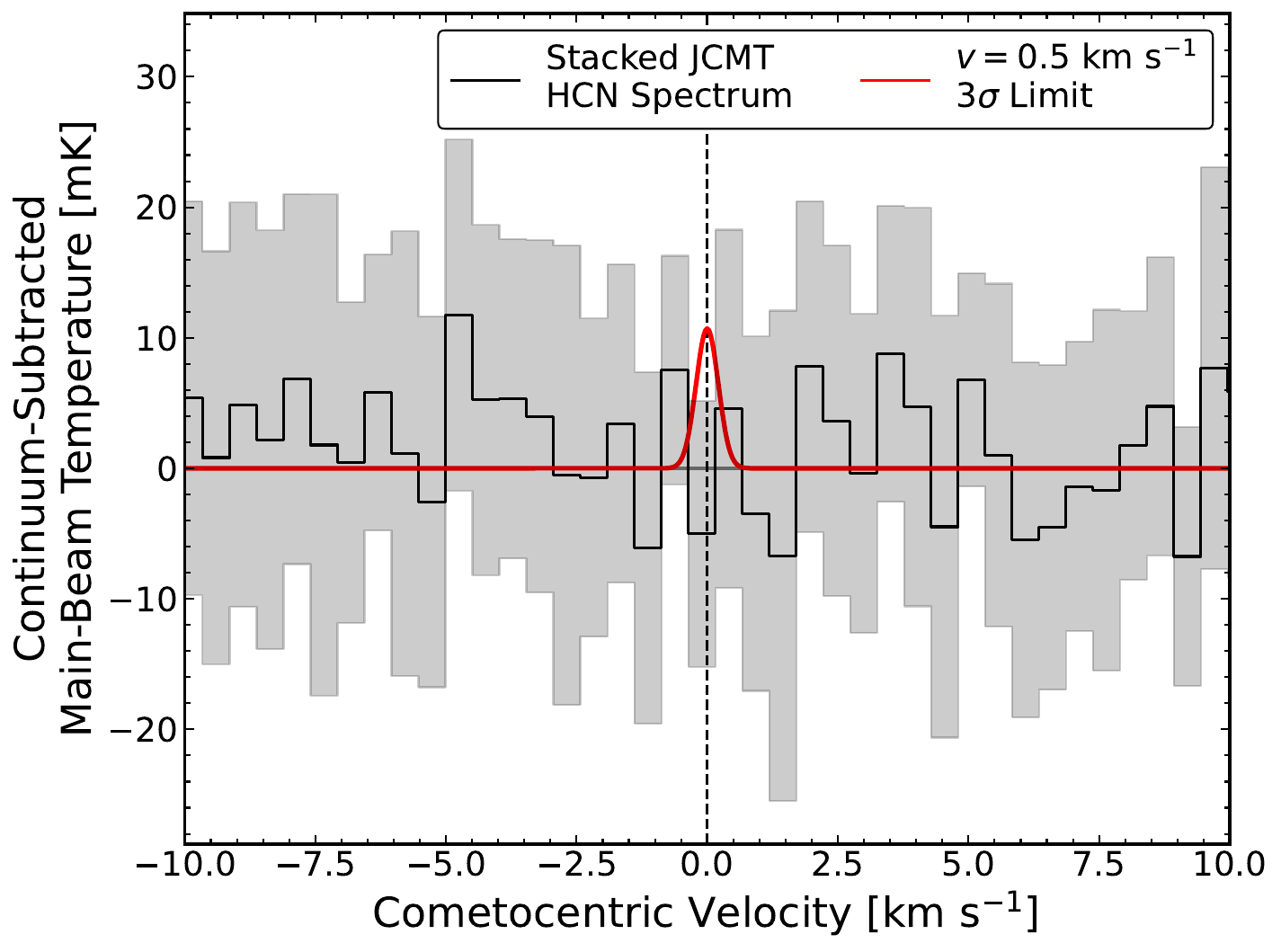}
 \caption{Stacked HCN($J = 3 - 2$) spectrum of 3I/ATLAS obtained with JCMT `\=U`\=u. \textit{Left:} Broadband spectrum shown in the cometocentric velocity frame at a binning of $1$ km s$^{-1}$ (black) and the scatter within each bin (gray). \textit{Right:} Spectrum centered on the expected line velocity at a binning of $0.5$ km s$^{-1}$ (black) and the scatter within each bin (gray). The red line indicates a Gaussian with a velocity FWHM of $0.5$ km s$^{-1}$, and an integrated intensity matching the 3$\sigma$ upper-limit on HCN emission. The vertical dashed lines mark zero velocity, and the horizontal lines mark the continuum level of zero K.}
 \label{fig:HCN_spectra}
\end{figure*}

Interstellar objects \citep[ISOs;][]{Jewitt2023ARAA, Fitzsimmons2024} are small bodies formed around other stars that briefly pass through the Solar System on hyperbolic orbits. While they are expected to be common \citep[e.g.,][]{Engelhardt2017, Trilling2017, Do2018}, we have detected just three thus far: 1I/`Oumuamua \citep{Meech2017}, 2I/Borisov \citep{borisov_2I_cbet}, and the recently-discovered 3I/ATLAS \citep{Seligman2025, Tonry2025}. As our sample of ISOs grows, they will provide a unique view of the population of small bodies within our Galaxy \citep[e.g.,][]{Do2018, ISSI_1I_review, Jewitt2023ARAA} and the physical and chemical conditions present during the formation of planetary systems around other stars \citep[e.g.,][]{Trilling2017, Seligman2022, Fitzsimmons2024}.

In Solar System comets, the sublimation of volatiles provides insight into the primordial conditions within the protosolar nebula \citep[e.g.,][]{Bocklee-Morvan2004}. Since interstellar comets likely form in a similar manner \citep[e.g.,][]{Seligman2022, Fitzsimmons2024}, they likewise trace the conditions present during the formation of their parent planetary systems. Of the three known ISOs, 1I/`Oumuamua is conspicuously inactive, with no observed coma and limits on the mass loss rate orders of magnitude below typical comets \citep[e.g.,][]{Meech2017, Jewitt2017, Ye2017, Trilling2018}. Conversely, 2I/Borisov had activity more similar to that of ordinary Solar System comets \citep[e.g.,][]{Fitzsimmons:2019, Jewitt2019b, Opitom:2019-borisov, Guzik:2020, Kareta2020, Aravind2021}. Its composition also resembled that of typical comets, with detections of molecular species like CN \cite{Fitzsimmons:2019}, C$_2$ \citep{Lin2020}, C$_3$ \citep{Lin2020}, NH$_2$ \citep{Bannister2020}, and OH \citep{Xing2020} and atomic species like [O I] \citep{McKay2020}, Ni \citep{Guzik2021}, and Fe \cite{Opitom2021}. However, the CO abundance was much higher than typical comets \citep{Bodewits2020, Cordiner2020}, likely indicating that it formed beyond the CO snow line in its parent protoplanetary disk.

3I/ATLAS also shows cometary activity \citep[e.g.,][]{Seligman2025, Jewitt2025, Opitom2025, Chandler2025, delaFuenteMarcos2025, Bolin2025}, possibly as far out at $\sim$6.7 au \citep{Tonry2025, Feinstein2025}. As 3I/ATLAS approached the Sun, several gas species were detected, including CN \citep{Rahatgaonkar2025}, Ni \citep{Rahatgaonkar2025}, Fe \citep{Hutsemekers25}, CO$_2$ \citep{Lisse2025, Cordiner2025}, OH \citep{Xing2025}, CO \citep{Cordiner2025}, and HCN \citep{Coulson2025}. Deep upper-limits on the presence of C$_2$ and C$_3$ indicate strong carbon-chain depletion relative to Solar System comets and 2I/Borisov \citep{Rahatgaonkar2025, SalazarManzano2025}, which itself is carbon-chain depleted. 

While many daughter molecules, such as CN, C$_2$, C$_3$, NH, NH$_2$, and OH are accessible through optical observations \citep[e.g.,][]{A'Hearn:1995, 2024come.book..459B}, observations at longer wavelengths, such as the infrared and sub-mm, provide access to parent molecules which are released into the coma directly from the nucleus. These include the major compounds dominating the coma composition, CO$_2$, CO, and H$_2$O, as well as less abundant molecules like HCN, CH$_3$OH, H$_2$CO, NH$_3$, and CS \citep[e.g.,][]{2024come.book..459B}. Broad wavelength coverage of these gas emissions in ISOs allows for detailed comparisons to Solar System comets \citep[e.g.,][]{Cordiner2020, Opitom2021} and insight into the physical mechanisms involved in driving their activity \citep[e.g.,][]{Rahatgaonkar2025, Hoogendam25_KCWI}.

Here we present the earliest constraints on HCN and CO emission from 3I/ATLAS based on observations with the James Clerk Maxwell Telescope (JCMT). As discussed in Section \ref{sec:data}, these observations took place in mid-July 2025, when 3I/ATLAS was at a heliocentric distance of $r_h \approx 3.9$ au. Combined with more recent detections of HCN from \citet{Coulson2025}, we examine the temporal evolution of the HCN emission in Section \ref{sec:temporal} and place this analysis in context with the evolution measured for other volatile species. Finally, we summarize our results in Section \ref{sec:summary}.

\section{Observations and Data}
\label{sec:data}

\subsection{Overview of JCMT Observations}

\begin{figure*}[t]
\centering
 \includegraphics[height=0.37\textwidth]{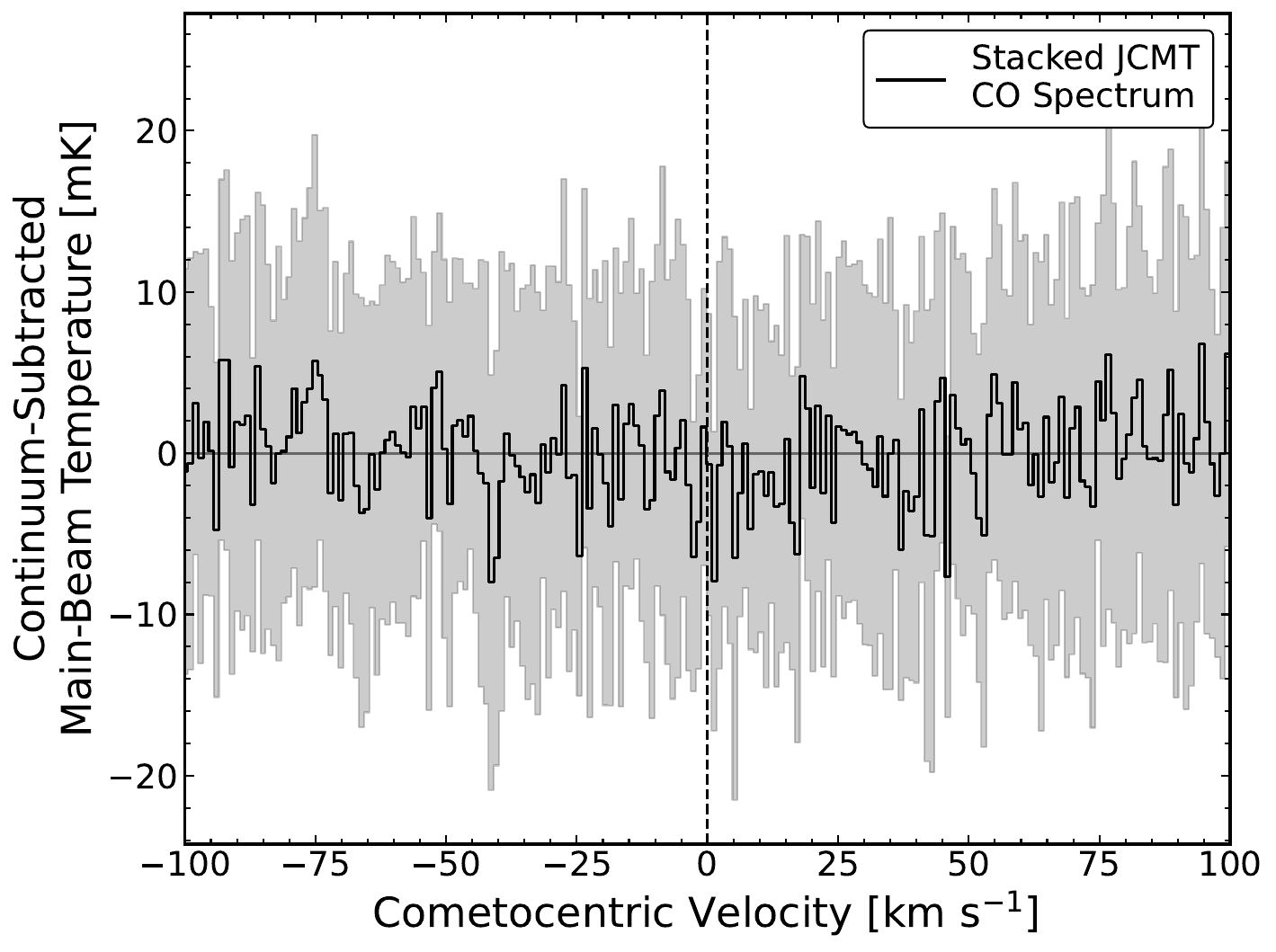}
 \includegraphics[height=0.37\textwidth]{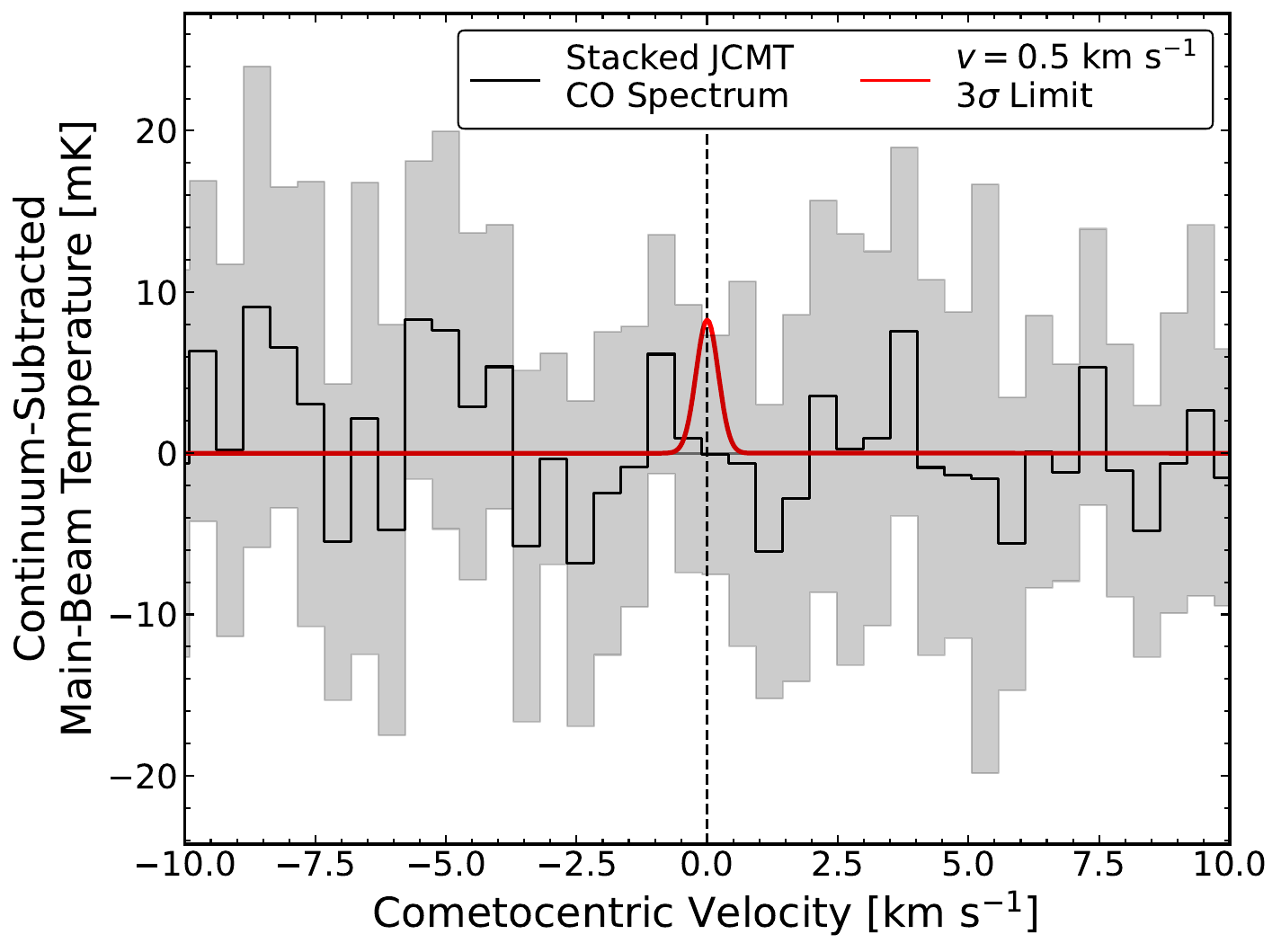}
 \caption{Stacked CO($J = 2 - 1$) spectrum of 3I/ATLAS obtained with JCMT `\=U`\=u. \textit{Left:} Broadband spectrum shown in the cometocentric velocity frame at a binning of $1$ km s$^{-1}$ (black) and the scatter within each bin (gray). \textit{Right:} Spectrum centered on the expected line velocity at a binning of $0.5$ km s$^{-1}$ (black) and the scatter within each bin (gray). The red line indicates a Gaussian with a velocity FWHM of $0.5$ km s$^{-1}$, and an integrated intensity matching the 3$\sigma$ upper-limit on CO emission. The vertical dashed lines mark zero velocity, and the horizontal lines mark the continuum level of zero K.}
 \label{fig:CO_spectra}
\end{figure*}

We observed the interstellar comet 3I/ATLAS with JCMT from 16 July 2025 through 21 July 2025 (UT) as part of program M25AH18B (PI:\ Hinkle). Our observations used the `\=U`\=u heterodyne receiver insert on the N\=amakanui instrument and the ACSIS (Auto Correlation Spectral Imaging System) backend. We used ACSIS in a configuration with a bandpass of 250 MHz and a spectral resolution of $\sim$37 kHz, optimal for observations of the narrow lines we targeted. Given the small ($<1$\arcsec) uncertainty on the ephemeris of 3I/ATLAS relative to the `\=U`\=u beamwidth at 230-265 GHz ($\approx20-18$\arcsec), we used stare mode to integrate on a single position (i.e., a $1 \times 1$ grid) and employed a beam-switch throw of 180\arcsec \ in azimuth. 

To limit Doppler smearing, we restricted the individual observations to 30 minutes. We checked the pointing and focus roughly hourly using strong sources close to 3I/ATLAS at the time of observation (e.g., RAFGL 1922) and took observations of flux standards (e.g., IRAS 16293-2422) typically before and after our science observations for calibration. Data reduction was performed using the \texttt{STARLINK} software \citep{Currie2014} and each individual scan was calibrated to the cometocentric frame using velocity information from the JPL Horizons ephemeris.

\subsection{HCN($J = 3 - 2$) Measurements}

On the first two nights, 16 July 2025 and 17 July 2025, we observed the $J = 3 - 2$ transition of HCN at 265.886 GHz. These observations spanned heliocentric distances of 4.01 to 3.97 au and geocentric distances of 3.10 to 3.07 au for 3I/ATLAS. The mean observation time of the individual scans was 60872.9 (MJD), with a full range of 1.2 days. Based on data collected by the JCMT Water Vapor Meter (WVM), which records the atmospheric opacity by intercepting part of the JCMT beam, our HCN spectra were obtained in good to moderate conditions with 225 GHz opacities between 0.06 and 0.11. The $T_{sys}$ range for our HCN observations is between 138 K and 199 K with a mean value of 151 K.

To obtain the final HCN spectrum, we co-added the Doppler-corrected spectra from all individual scans. This yielded a final stacked spectrum with a total integration time of 7.6 hours, corresponding to an effective on-source time of approximately 3.8 hours. There is no HCN emission line detected in our stacked spectrum, and we calculate an RMS scatter of 14 mK. For a line width of 0.5 km s$^{-1}$ \citep[e.g.,][]{Bocklee-Morvan2004, Cordiner2020} and the raw bin width of 0.034 km s$^{-1}$, this corresponds to a 3$\sigma$ upper-limit on the integrated line flux of $<$5.7 mK~km~s$^{-1}$, following the procedure of \citet{leonard01} and \citet{leonard07}.

The left panel of Figure \ref{fig:HCN_spectra} shows our continuum-subtracted stacked HCN spectrum binned at 1 km s$^{-1}$ over a wide velocity range. In the right panel of Figure \ref{fig:HCN_spectra}, we highlight velocities within 10 km s$^{-1}$ of the expected HCN line and at a binning of 0.5 km s$^{-1}$. We additionally use a Gaussian to illustrate the line profile corresponding to a feature with a velocity full-width at half-maximum (FWHM) 0.5 km s$^{-1}$ and the same integrated intensity as implied by our calculated RMS scatter.

We calculate an upper limit on the HCN gas production rate following \citet{Bocklee-Morvan2004} and \citet{Drahus2017}. Radio observations of comets indicate coma gas temperatures of a few tens of kelvin at $r_h \sim 4$~au (e.g., \citealt{Biver2002,Bockelee-Morvan2010}). The derived production rates depend only weakly on the assumed rotational temperature, varying by less than a factor of two over the plausible range ($\sim$20--40~K). For the intensity limits noted above, we derive a 3$\sigma$ upper-limit of $Q(HCN) < 1.7 \times 10^{24}$ s$^{-1}$ at $r_h = 4.01\mbox{ -- }3.97$ au, assuming a rotational temperature of 30 K and velocity FWHM of 0.5~km s$^{-1}$. This is both the earliest and the deepest limit on HCN gas production for 3I/ATLAS \citep[c.f.,][]{Coulson2025}.

\subsection{CO($J = 2 - 1$) Measurements}

For the following four nights, 18 July 2025 through 21 July 2025, we observed the $J = 2 - 1$ transition of CO at 230.538 GHz. These observations spanned heliocentric distances of 3.94 to 3.84 au and geocentric distances of 3.05 to 2.99 au for 3I/ATLAS. The mean observation time of the individual scans was 60875.9 (MJD), with a full range of 3.2 days. The JCMT WVM indicates that CO spectra were taken in a range of moderate to poor conditions. Most of the scans were taken with 225 GHz opacities between 0.07 and 0.13, but a handful of scans reached $\tau$ as high as 0.18.

To obtain our final CO spectrum, we co-added each of the Doppler-corrected spectra from the individual scans. This yielded a final stacked spectrum with a total integration time of 11.7 hours, corresponding to an effective on-source time of approximately 5.9 hours. There is no CO emission line detected in our stacked spectrum, and we calculate an RMS scatter of 11 mK. For a line width of 0.5 km s$^{-1}$ and the raw bin width of 0.039 km s$^{-1}$, this corresponds to a 3$\sigma$ upper-limit on the integrated line flux of $<$4.4 mK km s$^{-1}$. 

The left panel of Figure \ref{fig:CO_spectra} shows our continuum-subtracted stacked CO spectrum binned at 1 km s$^{-1}$ over a wide velocity range. In the right panel of Figure \ref{fig:CO_spectra}, we highlight velocities within 10 km s$^{-1}$ of the expected CO line and at a binning of 0.5 km s$^{-1}$. Using a Gaussian, we illustrate the line profile corresponding to a feature with a velocity FWHM of 0.5 km s$^{-1}$ and the same integrated intensity as implied by our calculated RMS scatter.

Following the same procedure as for HCN, we calculate a 3$\sigma$ upper-limit on the CO gas production rate of $Q(CO) < 1.1 \times 10^{27}$ s$^{-1}$ at $r_h = 3.94\mbox{ -- }3.84$ au. This represents the earliest limit on CO gas production for 3I/ATLAS. However, it is less stringent than the limit of $Q(CO) < 2.8 \times 10^{26}$ s$^{-1}$ at $r_h = 3.2$ au from SPHEREx data \citep{Lisse2025} and the detection of $Q(CO) = (1.70 \pm 0.04) \times 10^{26}$ s$^{-1}$ at $r_h = 3.32$ au from JWST NIRSpec data \citep{Cordiner2025}.

\begin{figure*}
\centering
 \includegraphics[width=0.48\textwidth]{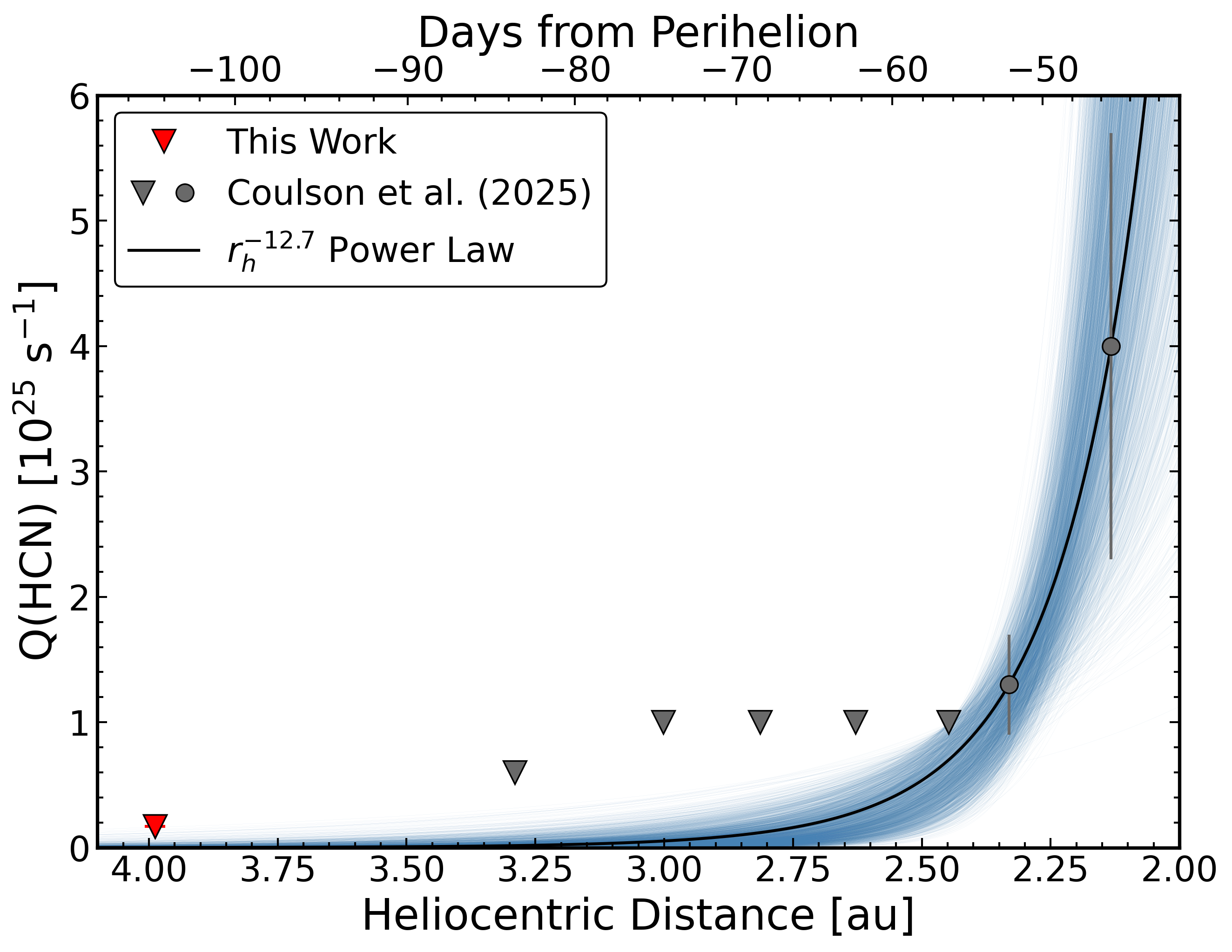}
 \includegraphics[width=0.48\textwidth]{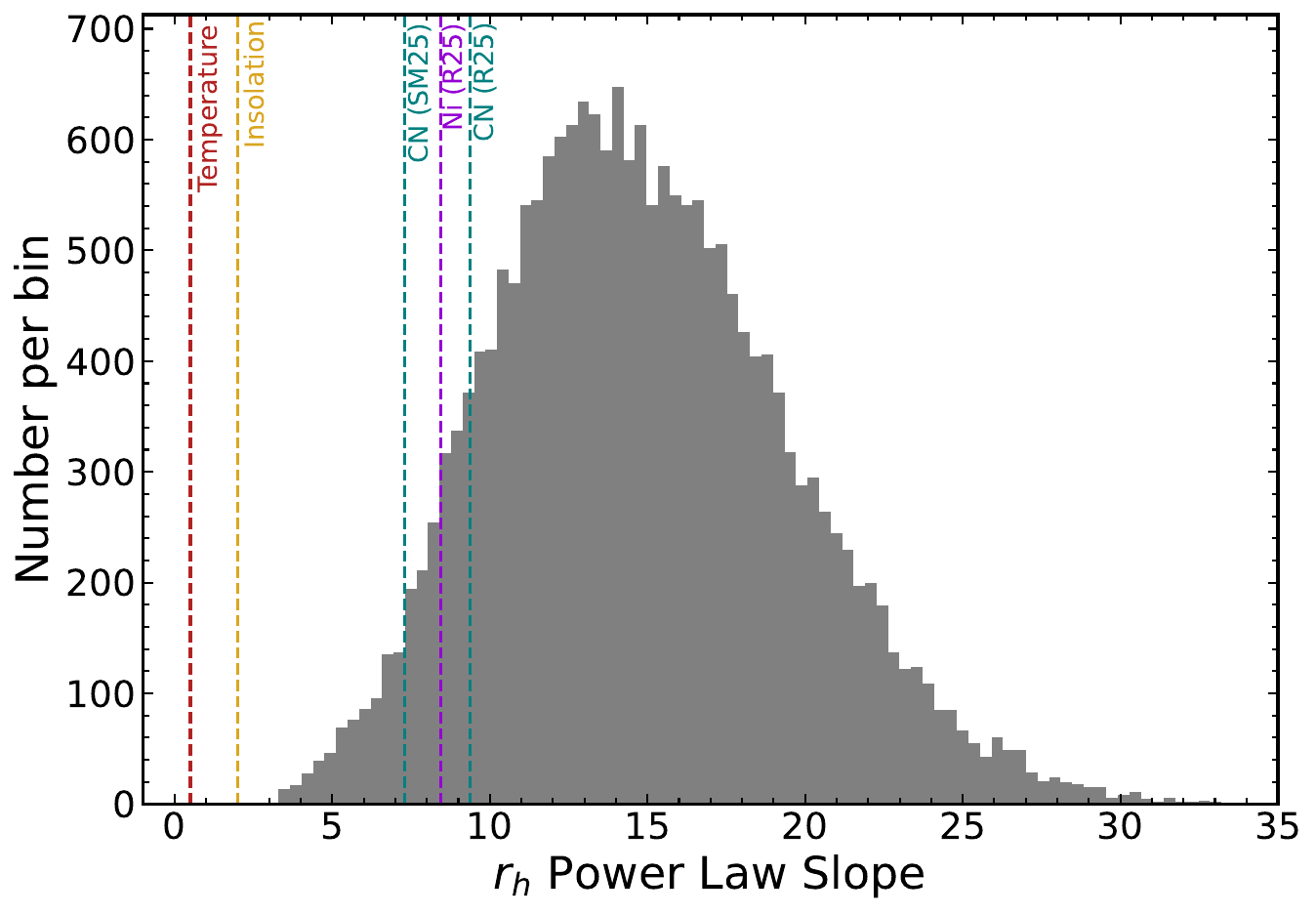}
 \caption{\textit{Left:} Temporal evolution of HCN($J = 3 - 2$) emission from 3I/ATLAS. Our constraint is shown in red, and data from \citet{Coulson2025} are shown in gray. Downward-facing triangles indicate 3$\sigma$ upper limits, and circles are detections with corresponding 1$\sigma$ uncertainties. The solid black line is the best-fit power-law model, and the blue lines are 3000 random realizations of allowed Monte Carlo fits. \textit{Right:} Histogram of power-law slopes resulting from 20,000 Monte Carlo iterations. Dashed vertical lines show typical slopes such as the $r_h^{-0.5}$ temperature scaling (red), the $r_h^{-2}$ scaling from Solar insolation (gold), and measured dependences of CN (teal) and Ni (purple) in 3I/ATLAS from \citet[][R25]{Rahatgaonkar2025} and \citet[][SM25]{SalazarManzano2025}.}
 \label{fig:HCN_fits}
\end{figure*}

\section{Temporal Evolution of HCN in 3I/ATLAS}
\label{sec:temporal}

The left panel of Figure \ref{fig:HCN_fits} shows the temporal evolution of HCN gas production rates from 3I/ATLAS, comprising our measurements and those of \citet{Coulson2025}. These data span heliocentric distances of 3.99 to 2.13 au, corresponding to $-$104.6 to $-$45.2 days from perihelion. While many of the early HCN constraints are upper limits, they still provide valuable constraints on when cometary activity turned on in 3I/ATLAS. These are important for understanding the context with other gaseous species such as CN (see Section \ref{sec:CN}), and constrain the production rate dependence on heliocentric distance (see Section \ref{sec:HCN_rates}).

\subsection{Comparison with CN}
\label{sec:CN}

Optical observations have indicated the presence of CN, a plausible daughter species of HCN \citep{Fray2005}, in the coma of 3I/ATLAS \citep[][]{Rahatgaonkar2025, SalazarManzano2025, Schleicher2025, Hoogendam25_KCWI, Hoogendam25_SNIFS}. A weak detection of CN occurred as early as 2025 August 10 \citep{SalazarManzano2025}, when the comet was at $r_h = 3.16$ au. Subsequent spectra of 3I/ATLAS have shown progressively stronger CN emission \citep[][]{SalazarManzano2025, Rahatgaonkar2025, Hoogendam25_SNIFS}. Although the general trend of increasing CN production as 3I/ATLAS nears the Sun is clear, there is a large scatter in the estimated gas production rates. For instance, at $r_h \approx 3$ au, estimates of $Q(CN)$ vary widely from modest values of $Q(CN) = (2.3 \pm 0.5) \times 10^{23}$ s$^{-1}$ \citep{Rahatgaonkar2025} and $Q(CN) = (8.4 \pm 3.2) \times 10^{23}$ s$^{-1}$ \citep[][]{Hoogendam25_SNIFS} to values roughly an order of magnitude higher at $Q(CN) = (6.3 \pm 0.3) \times 10^{24}$ s$^{-1}$ \citep{SalazarManzano2025}. The CN production rates for 3I/ATLAS are similar to those estimated for 2I/Borisov at comparable heliocentric distances \citep[e.g.,][$\approx$$(1\mbox{ -- }4) \times 10^{24}$ s$^{-1}$]{Opitom:2019-borisov, Fitzsimmons:2019, Kareta2020} and on the low end of the range typically seen for Solar System comets \citep[e.g.,][$\approx$$10^{24}\mbox{ -- }10^{26}$ s$^{-1}$]{A'Hearn:1995, mumma11}.

Regardless, at these heliocentric distances, the CN gas production rates as compared to the HCN upper-limits remain consistent with CN production through HCN photolysis \citep{Coulson2025}. The evolution of CN, particularly at close heliocentric distances, is poorly constrained relative to HCN. The closest published pre-perihelion data was taken when 3I/ATLAS was at a heliocentric distance of 2.47 au at a value of $Q(CN) = (3.1 \pm 0.5) \times 10^{24}$ s$^{-1}$ \citep{Hoogendam25_SNIFS}. These data and extrapolations to distances at which HCN is detected \citep{Coulson2025} suggest that the $Q(CN)/Q(HCN)$ ratios stay close to or below unity as the HCN emission from 3I/ATLAS begins to turn on.

\subsection{Production Rate Dependence on $r_h$}
\label{sec:HCN_rates}

Several molecular and atomic species have now been detected in 3I/ATLAS in multiple epochs, including CN \citep[][]{Rahatgaonkar2025, SalazarManzano2025, Schleicher2025, Hoogendam25_KCWI, Hoogendam25_SNIFS, Hoogendam26_KCWI}, Ni \citep[][]{Rahatgaonkar2025, Hoogendam25_KCWI, Hoogendam25_SNIFS, Hoogendam26_KCWI}, Fe \citep{Hutsemekers25, Hoogendam26_KCWI}, and HCN \citep{Coulson2025}. Thus far, the gas production rate dependence on heliocentric distance has been investigated for CN and Ni. For CN, \citet{Rahatgaonkar2025} find $Q(CN) \propto r_h^{-9.38 \pm 1.20}$ and \citet{SalazarManzano2025} obtain a consistent fit of $Q(CN) \propto r_h^{-7.3 \pm 1.1}$, both much steeper than typical Solar System comets \citep[e.g.,][]{A'Hearn:1995, Manfroid2021} and the relatively flat evolution of 2I/Borisov \citep{Opitom:2019-borisov}. For Ni, \citet{Rahatgaonkar2025} find $Q(Ni) \propto r_h^{-8.43 \pm 0.79}$, also steep compared to Solar System comets \citep{Manfroid2021}.

Here we compute the dependence of $Q(HCN)$ on heliocentric distance, combining our measurements with those of \citet{Coulson2025}. Adopting a model of the form $Q(HCN) = Ar_h^{-n}$, accounting for measured upper-limits, and fitting in log space with \texttt{\detokenize{scipy.optimize.curve_fit}} and the Trust Region Reflective algorithm, we find a best-fit slope of $n = 12.7$. With only two HCN detections, the power-law fits are poorly constrained, but we nevertheless confirmed that the best fit was not sensitive to our initial guess for the minimization. Alternatively, a model of HCN production turning on suddenly and evolving linearly thereafter yields a slope of $-1.4 \times 10^{26}$ s$^{-1}$ au$^{-1}$ beginning at $r_h = 2.43$ au. Given the detection of other volatiles beyond this distance, we elect to investigate the power-law model further.

We estimate the uncertainties on this slope through Monte Carlo sampling, perturbing the measured HCN detections assuming Gaussian errors and running 20,000 iterations of the power-law fit, again accounting for the upper-limits. This yields a power-law slope of $n = 12.7^{+6.9}_{-2.5}$. This fit is consistent with measurements of HCN($J = 4 - 3$) taken with the Atacama Compact Array \citep{roth25}, particularly after accounting for differences in the line studied and modeling assumptions. A histogram illustrating the results of these Monte Carlo iterations is shown in the right panel of Figure \ref{fig:HCN_fits}. Consistent with the behavior illustrated in the left panel of Figure \ref{fig:HCN_fits}, although most fits have a high power-law index, there are a small number of significantly shallower fits that can also describe the data. Our upper-limit of $Q(HCN) < 1.7 \times 10^{24}$ s$^{-1}$ at $r_h \simeq 3.99$ au provides an important constraint in ruling out the shallowest models.

Despite the large uncertainties, it is clear that the power-law dependence of the HCN gas production rate on heliocentric distance for 3I/ATLAS is steep. Both our new upper-limit and the closest upper limit on HCN emission from \citet{Coulson2025} provide key constraints, requiring that the slope be steep enough to produce the later detections at $r_h \leq 2.33$ au, while not overproducing HCN slightly farther away at $r_h \approx 2.45$ au and significantly farther away at $r_h \gtrsim 4$ au. Furthermore, the steep slope for the HCN rate dependence is similar to the slopes found for Ni \citep{Rahatgaonkar2025} and CN \citep{Rahatgaonkar2025, SalazarManzano2025}. It is also steeper than the $r_h^{-2}$ to $r_h^{-4.5}$ dependences often seen for HCN emission in Solar System comets \citep[e.g.,][]{Biver1999, Biver2000, Biver2011, Wirstrom2016, Li2025}. These systematically higher rate dependences may indicate a difference in the formation of 3I/ATLAS relative to Solar System comets \citep[e.g.,][]{Jewitt2023ARAA, Fitzsimmons2024} or subsequent processing on its passage through the Galaxy \citep[e.g.,][]{Maggiolo2025, Ye2025}.

\section{Summary} \label{sec:summary}

We present the earliest constraints on HCN and CO emission from the interstellar comet 3I/ATLAS based on observations with `\=U`\=u on JCMT. The observations occured between 16 July 2025 and 21 July 2025 (UT), when 3I/ATLAS was at a heliocentric distance between 4.03 and 3.78 au. We additionally combine our measurements with HCN constraints from \citet{Coulson2025} to provide a more holistic view of the HCN evolution for 3I/ATLAS. Our results are summarized as follows:

\begin{itemize}

\item HCN is undetected in our stacked spectrum. We derive a 3$\sigma$ upper limit of $Q(HCN) < 1.7 \times 10^{24}$ s$^{-1}$ at $r_h = 4.01\mbox{ -- }3.97$ au. This represents the deepest limit on HCN emission from 3I/ATLAS to date.

\item CO is undetected in our stacked spectrum. We derive a 3$\sigma$ upper limit of $Q(CO) < 1.1 \times 10^{27}$ s$^{-1}$ at $r_h = 3.94\mbox{ -- }3.84$ au.

\item An extrapolation of CN gas production rates from optical spectroscopy indicates that the $Q(CN)/Q(HCN)$ ratios remain close to or below unity as HCN emission in 3I/ATLAS begins, consistent with CN production from HCN photolysis. \vspace{-0.1em}

\item The HCN gas production rate evolution with heliocentric distance is steep, with a power-law index of $n = 12.7^{+6.9}_{-2.5}$. Our $Q(HCN)$ upper-limit at $r_h \simeq 3.99$ au constrains power-laws with the shallowest slopes. This is consistent with rate dependence measurements for other compounds in 3I/ATLAS and steep relative to Solar System comets.

\end{itemize}

Pre-perihelion measurements of volatile activity from 3I/ATLAS indicate a much steeper evolution with heliocentric distance than is typical for Solar System comets. These differences likely encode information about its origin and environment preceding its sojourn through our Solar System. Continued observations of 3I/ATLAS post-perihelion will provide further insights into the physical and chemical properties of this interstellar interloper. Combined with the expected increase of interstellar objects discovered by the Legacy Survey of Space and Time \citep[LSST;][]{LSST_2019} on the Vera C.\ Rubin Observatory, these observations will shed light on the process of forming stellar and planetary systems in our Galaxy.

\section*{Acknowledgments}

We thank the referee for helpful comments and suggestions that
have improved the quality of this manuscript. We thank JCMT astronomers Harold Pe\~na and Jia-Wei Wang for their support in scheduling these observations, and JCMT telescope operators Andrea McCl\"oskey, Dan Bintley, and Ryan Torres for their work on conducting these observations.

J.T.H. acknowledges support from NASA through the NASA Hubble Fellowship grant HST-HF2-51577.001-A, awarded by STScI. STScI is operated by the Association of Universities for Research in Astronomy, Incorporated, under NASA contract NAS5-26555.

K.J.M., J.J.W., and A.H.\ acknowledge support from the Simons Foundation through SFI-PD-Pivot Mentor-00009672. W.B.H. acknowledges support from the NSF Graduate Research Fellowship Program under Grant No. 2236415. The Shappee group at the University of Hawai`i is supported with funds from NSF (grant AST-2407205) and NASA (grants HST-GO-17087, 80NSSC24K0521, 80NSSC24K0490, 80NSSC23K1431).

These observations were obtained by the James Clerk Maxwell Telescope, operated by the East Asian Observatory on behalf of Academia Sinica Institute of Astronomy and Astrophysics and the National Astronomical Research Institute of Thailand. Additional funding support is provided by the Science and Technology Facilities Council of the United Kingdom and participating universities and organizations in the United Kingdom and Canada. \\

\noindent \textit{Facilities:} JCMT \\
\textit{Software:} \texttt{STARLINK}

\bibliography{bibliography}{}
\bibliographystyle{aasjournal}

\end{document}